\documentclass[12pt]{article}
\usepackage{epsfig}

\newcommand{\bea}{\begin{eqnarray}}
\newcommand{\eea}{\end{eqnarray}}
\newcommand{\eL}{\epsilon_L}
\newcommand{\eLv}{\epsilon_L^{v}}
\newcommand{\eLc}{\epsilon_L^{c}}
\newcommand{\eR}{\epsilon_R}
\newcommand{\eS}{\epsilon_S}
\newcommand{\eT}{\epsilon_T}
\newcommand{\eP}{\epsilon_P}
\newcommand{\teL}{\tilde{\epsilon}_L}
\newcommand{\teR}{\tilde{\epsilon}_R}
\newcommand{\teS}{\tilde{\epsilon}_S}
\newcommand{\teT}{\tilde{\epsilon}_T}
\newcommand{\teP}{\tilde{\epsilon}_P}

\def\beq{\begin{equation}}
\def\eeq#1{\label{#1}\end{equation}}
\def\eeqn{\end{equation}}

\def\beqa{\begin{eqnarray}}
\def\eeqa#1{\label{#1}\end{eqnarray}}
\def\eeqan{\end{eqnarray}}

\let\bar=\overbar

\def\Dslash{\not{\hbox{\kern-4pt $D$}}}
\def\dslash{\not{\hbox{\kern-2pt $\del$}}}

\def\msb{{\bar{\ssstyle M \kern -1pt S}}}

\def\Title#1{\begin{center} {\Large {\bf #1} } \end{center}}
\def\Author#1{\begin{center}{ \sc #1} \end{center}}
\def\Address#1{\begin{center}{ \it #1} \end{center}}

\begin{document}

\Title{\bf Probing non-standard charged-current interactions: from cold neutrons to the LHC}
\Author{Mart\'in Gonz\'alez-Alonso\index{Gonz\'alez-Alonso, M.}}
\Address{Department of Physics, University of Wisconsin, Madison, WI 53706, USA}


\vspace{0.1cm}
\begin{quotation}

It is well known that semileptonic decays of light hadrons and nuclei can be used not only to determine the CKM element $V_{ud}$ with high accuracy, but also as probes of physics beyond the Standard Model. In this talk I review recent works that studied this within an Effective Field Theory framework, comparing the sensitivity of different low-energy and LHC observables. A clear complementarity between low- and high-energy searches it is found.

\end{quotation}

\begin{center}
Proceedings of CKM 2012,\\
the 7th International Workshop on the CKM Unitarity Triangle,\\
University of Cincinnati, USA, 28 September - 2 October 2012      
\end{center}
\vspace{0.3cm}


Semileptonic decays of hadrons have been used very successfully to determine different CKM matrix elements during the last decades. In some cases both the experimental measurements and the Standard Model predictions are extremely precise, making the possibility of probing New Physics (NP) a very interesting and legitimate question. It is obvious that any physical observable is sensitive to New Physics at some level, but the real question is if these measurements of (semi)leptonic decays of hadrons have a higher sensitivity than the rest of observables for some kind of exotic interaction. In this talk we will review the work done in Refs.~\cite{Cirigliano:2009wk,Bhattacharya:2011qm,Cirigliano:2012ab}, where this question was addressed within an Effective Field Theory approach and focusing on the case of light hadrons and nuclei.

The $V_{ud}$ matrix element can be determined from nuclear, neutron or pion 
decays with different levels of accuracy, as nicely reviewed in several talks in this conference \cite{CKM2012}. Here we are interested in the potential of these measurements to probe New Physics. Even if we are forced to deal with hadrons at the experiment, it is worth going one step down in the theoretical description and using a Lagrangian where the quarks are the active degrees of freedom, since all these decays probe the same partonic process. In this way all the NP bounds can be casted in a common language and one can compare them and evaluate their interplay.

For this reason we will work with the following low-scale  $O(1 \ {\rm GeV})$  effective Lagrangian for semileptonic transitions:
\bea
{\cal L}_{\rm CC}  &=&
- \frac{G_F V_{ud}}{\sqrt{2}} \  \Big[ \ \Big( 1 +  \eL \Big) \  
\bar{e}  \gamma_\mu  (1 - \gamma_5)   \nu_{\ell}  \cdot \bar{u}   \gamma^\mu  (1 - \gamma_5)  d   \\
&+& \teL  \ \ \bar{e}  \gamma_\mu  (1 + \gamma_5)   \nu_{\ell}  \cdot \bar{u}   \gamma^\mu  (1 - \gamma_5)  d  \nonumber\\
&+&   \eR   \  \   \bar{e}  \gamma_\mu  (1 - \gamma_5)   \nu_{\ell}  \cdot \bar{u}   \gamma^\mu  (1 + \gamma_5)  d  
\ + \  \tilde{ \epsilon}_R   \  \   \bar{e}  \gamma_\mu  (1 +  \gamma_5)   \nu_{\ell} \cdot \bar{u}   \gamma^\mu  (1 + \gamma_5)  d  \nonumber\\
&+&  \eT   \   \bar{e}   \sigma_{\mu \nu} (1 - \gamma_5) \nu_{\ell}    \cdot  \bar{u}   \sigma^{\mu \nu} (1 - \gamma_5) d
\ + \  \teT      \   \bar{e}   \sigma_{\mu \nu} (1 + \gamma_5) \nu_{\ell}    \cdot  \bar{u}   \sigma^{\mu \nu} (1 + \gamma_5) d  \nonumber \\
&+&  \eS  \  \  \bar{e}  (1 - \gamma_5) \nu_{\ell}  \cdot  \bar{u} d  \ + \  \teS  \  \  \bar{e}  (1 +  \gamma_5) \nu_{\ell}  \cdot  \bar{u} d  \nonumber \\
&-& \eP  \  \   \bar{e}  (1 - \gamma_5) \nu_{\ell}  \cdot  \bar{u} \gamma_5 d 
\ - \  \teP  \  \   \bar{e}  (1 + \gamma_5) \nu_{\ell}  \cdot  \bar{u} \gamma_5 d  \ \Big]+{\rm h.c.}~. \nonumber
\eea
The non-standard couplings $\epsilon_i$ and $\tilde{\epsilon}_i$ are functions of the masses and couplings of the new heavy particles yet to be discovered, in the same way as the Fermi constant is a function of the W mass and the weak coupling. However we do not need to know these functions to compare the NP reach of different beta decay experiments.   

It is worth emphasizing that the extraction of bounds on the NP couplings $\epsilon$ requires the calculation of the hadronic form factors associated to each individual process. Schematically we have $C = \rm{FF} \times \epsilon$, where $C$ are the hadronic level coefficients that one actually probes experimentally and FF are the form factors. It is trivial to see that an extremely precise measurement will not set strong bounds on a certain NP coupling $\epsilon$ unless we can somehow put a meaningful lower bound on the absolute value of the corresponding form factor. Therefore the NP bounds depend both on the experimental accuracy and our ability to calculate these form factors that sometimes are not very well known, not only because QCD is complicated but also because they are not generated in the Standard Model, and thus they have not received so much attention historically.

Table~\ref{tab:summary} shows a summary of the current bounds on these ten parameters $\epsilon_i$ and $\tilde{\epsilon}_i$ that encode all the NP impact on beta decays. The pseudo-scalar couplings are somehow special and are strongly constrained by the ratio $\rm{R_\pi = \Gamma(\pi\to e\nu) /\Gamma(\pi\to \mu\nu)}$ \cite{Britton:1992pg} due to the helicity enhancement \footnote{Although not shown in Table~\ref{tab:summary}, the ratio $R_\pi$ is also a very powerful probe of scalar and tensor interactions since they generate radiatively a pseudo-scalar interaction \cite{Campbell:2003ir}. See Refs.~\cite{Bhattacharya:2011qm,Cirigliano:2012ab} for more details.}. The strongest constraints on (axial)vector and tensor couplings, $\epsilon_{L,R}$ and $\eT$, come from CKM unitarity tests \cite{Cirigliano:2009wk} and radiative pion decay \cite{Bychkov:2008ws} respectively, whereas for the rest of couplings the most stringent limits come from nuclear beta decays \cite{Severijns:2006dr,Hardy:2004id}. The $\tilde{\epsilon}_i$ coefficients affect the observables only quadratically, since they come with a right-handed (RH) neutrino, whereas the $\epsilon_i$ terms can interfere with the Standard Model (SM), and thus low-energy experiments are more sensitive to them.  More details about these different bounds can be found in Refs.~\cite{Bhattacharya:2011qm,Cirigliano:2012ab}. 

Future measurements with (ultra)colds will measure the Fierz term $b$ of neutron beta decay with an expected sensitivity of $10^{-3}$ \cite{Pocanic:2008pu,UCNb}, what will improve the current constraints on the tensor coupling $\eT$ \cite{Bhattacharya:2011qm,Jackson1957zz}. In order to improve the constraints on the scalar one $\eS$ we will need one order of magnitude more of sensitivity in the $b$ measurements.

After this brief review of current and future low-energy bounds on these non-standard couplings, we move on now to high-energy probes and specifically LHC searches. The obvious channel to study is $pp\to e + MET$, since the underlying partonic process is the same ($\bar{u}d\to e\bar{\nu}$), and so one expects that whatever NP is generating non-zero $\epsilon_i$ coefficients will also modify the cross-section of this process at some level. 

Assuming that the heavy mediators that generate these non-standard couplings are too massive to be produced at the LHC we can follow again an EFT approach where the heavy particles have been integrated. It is well-known that the first corrections to the SM Lagrangian are given by dimension six-operators ${\cal O}_i$ 
 that will have a certain impact in our observable:
\bea
{\cal L}_{eff} = \sum_i \frac{\alpha_i}{\Lambda^2} {\cal O}_i ~~~\to ~~~\sigma_{pp\to e+MET} = f \left( \frac{\alpha_i}{\Lambda^2} \right)
\eea 
where $\alpha_i$ are the Wilson coefficients. Matching the high- and low-energy effective Lagrangians one can derive the relations $\alpha_i = \alpha_i (\epsilon_j)$ (see Refs.~\cite{Bhattacharya:2011qm,Cirigliano:2012ab}) and in this way we can put bounds on the $\epsilon_i$ coefficients from the LHC searches. Doing so the cross-section with transverse mass higher than $\overline{m}_T$ takes the following form:
\bea
\label{eq:sigmamt}
\sigma_{pp\to e\nu}(m_T \!\!>\! \overline{m}_T) &=&    
\sigma_W \Big[ ( 1 +   \eLv)^2+  |\teL|^2  + | \eR|^2 \Big] - 2 \, \sigma_{WL}\,  \eLc \left( 1 +  \eLv \right) \\
&&\hspace{-2.1cm}
+~ \sigma_R  \Big[ |\teR|^2 \!+ |\eLc|^2 \Big] + \sigma_S \Big[ |\eS|^2  \!+  |\teS|^2  \!+ |\eP|^2 \!+  |\teP|^2   \Big] +  \sigma_T  \Big[ |\eT|^2  \!+  |\teT|^2     \Big]~,\nonumber 
\eea 
where $\sigma_W (\bar{m}_T)$ is the SM contribution and $\sigma_{WL, R, S,T}(\bar{m}_T)$ are new functions, which explicit form can be found in Ref.~\cite{Cirigliano:2012ab}. The key point here is that these functions are several orders of magnitudes larger than the SM contribution, what compensates for the smallness of the NP coefficients and makes possible to set significant bounds on them from these searches. For the same reason this observable is not very sensitive to $\eLv$, $\teL$ and $\eR$.    Moreover it is worth imposing a lower limit on the transverse mass of the lepton pair $m_T$ in order to increase the signal/background ratio.

So far both CMS and ATLAS results are in good agreement with the SM prediction, what sets limits on the different NP couplings. Table~\ref{tab:summary} shows the bounds obtained with 5 fb$^{-1}$ of data recorded at $\sqrt{s}=$ 7 TeV by the CMS Collaboration in the $pp\to e+MET$ channel \cite{ENcms5fb}, after imposing a cut of 1.2 TeV over $m_T$. We refer the reader to Refs.~\cite{Bhattacharya:2011qm,Cirigliano:2012ab} for more details. Needless to say, the LHC sensitivity to these non-standard couplings gets stronger as more data is collected. 

Comparing low- and high-energy bounds in Table~\ref{tab:summary} we see that both probes are needed to get a complete picture of non-standard charged current interactions.  For some couplings like the pseudoscalar ones low-energy probes are much more powerful, whereas for interactions involving RH neutrinos the LHC will dominate the search \cite{Cirigliano:2012ab}. In some other cases there is an interesting competition between low- and high-energy searches, like in the scalar and tensor cases.

In this talk we have focused on non-standard couplings involving only first-generation fermions, but this analysis can be applied to other flavor structures. For example Ref.~\cite{Cirigliano:2009wk} studied the limits on operators involving down or strange quarks, that would affect the $V_{ud}$ and $V_{us}$ determinations altering the CKM unitarity tests. Under certain assumptions about the flavor structure of the new operators, it was shown that such unitarity tests can complement collider searches, in another nice example of the interplay between the energy and the intensity frontiers.

\begin{table}
\centering
\begin{tabular}{|c|c|c|c|c|c|c|}
\hline
                  & $ |\epsilon_L^{(v)}| $           &  	 $ \epsilon_L^{(c)}$      &       $|\epsilon_R|$           &     $|\epsilon_P |$    &      $|\epsilon_S|$         &       $| \epsilon_T|$    \\
\hline 
\hline
Low energy	&   		0.05	                   &          0.05                    &               0.05                           &            0.06             &                 0.8                 &              0.1                 \\
\hline 
 LHC		$(e \nu)$   &   		-	                 &	        (-0.3,+0.8)	       &               -                   &               1.3                   &                1.3                       &             0.3               \\
\hline
				\multicolumn{7}{c}{}              \\
\hline
				&	\multicolumn{2}{c|}{$ |  \teL| $}			&	$| \teR| $           		&	$| \teP|$			&	$| \teS|$		&	$|\teT|$               \\
\hline 
\hline
Low energy	&		\multicolumn{2}{c|}{6}					&	6						&	0.03				&		14			&	3.0				\\
\hline 
LHC $(e \nu)$		&	\multicolumn{2}{c|}{-}					&	0.5						&	1.3					&		1.3			&	0.3				\\
\hline 
\end{tabular} 
\caption{Summary of 90\% CL bounds (in units of  $10^{-2}$)  
on the non-standard couplings $\epsilon_\alpha$ and $\tilde{\epsilon}_\alpha$ obtained  from low-energy and LHC searches. 
Notice that high-energy searches probe separately the vertex correction $\eLv$ and contact $\eLc$ contributions to the coupling $\eL$, defined in Ref.~\cite{Cirigliano:2012ab}.
\label{tab:summary}}
\end{table}

\bigskip
This work was supported by the U.S. DOE contract DE-FG02-08ER41531 and the Wisconsin Alumni Research Foundation.

\end{document}